\documentclass[12pt,preprint]{emulateapj}

\usepackage{natbib}
\usepackage{amsmath,amssymb,graphicx,soul,color}

\shorttitle{LAT observations of GRB 110709A, 111117A and 120107A}

\slugcomment{Draft version on April 30, 2012}

\begin{document}

\title{GRB 110709A, 111117A and 120107A: Faint high-energy gamma-ray photon emission from $Fermi$/LAT observations and demographic implications}

\author{ WeiKang~Zheng\altaffilmark{1}
\email{zwk@umich.edu}
Carl~W.~Akerlof\altaffilmark{1}, 
Shashi~B.~Pandey\altaffilmark{2}, 
Timothy~A.~McKay\altaffilmark{1}
BinBin~Zhang\altaffilmark{3},
Bing~Zhang\altaffilmark{4}, and
Takanori~Sakamoto\altaffilmark{5,6,7}
}

\altaffiltext{1} {Department of Physics, University of Michigan, 450 Church Street, Ann Arbor, MI, 48109, USA}
\altaffiltext{2} {Aryabhatta Research Institute of Observational Sciences, Manora Peak, Nainital, India, 263129}
\altaffiltext{3} {Department of Astronomy \& Astrophysics, Pennsylvania State University, University Park, PA 16802, USA}
\altaffiltext{4} {Department of Physics and Astronomy, University of Nevada, Las Vegas, NV, 89154, USA}
\altaffiltext{5}{Center for Research and Exploration in Space Science and Technology (CRESST), NASA Goddard Space Flight Center, Greenbelt, MD 20771}
\altaffiltext{6}{Joint Center for Astrophysics, University of Maryland, Baltimore County, 1000 Hilltop Circle, Baltimore, MD 21250}
\altaffiltext{7}{NASA Goddard Space Flight Center, Greenbelt, MD 20771}

\shortauthors{Zheng et al. 2012}

\begin{abstract}
Launched on June 11, 2008, the LAT instrument onboard the $Fermi$ Gamma-ray Space Telescope has provided a rare opportunity to study high energy photon emission from gamma-ray bursts. Although the majority of such events (27) have been identified by the Fermi LAT Collaboration, four were uncovered by using more sensitive statistical techniques (Akerlof et al 2010, Akerlof et al 2011, Zheng et al 2012). In this paper, we continue our earlier work by finding three more GRBs associated with high energy photon emission, GRB 110709A, 111117A and 120107A. To systematize our matched filter approach, a pipeline has been developed to identify these objects in near real time. GRB 120107A is the first product of this analysis procedure. Despite the reduced threshold for identification, the number of GRB events has not increased significantly. This relative dearth of events with low photon number prompted a study of the apparent photon number distribution. We find an extremely good fit to a simple power-law with an exponent of -1.8 $\pm$ 0.3 for the differential distribution. As might be expected, there is a substantial correlation between the number of lower energy photons detected by the GBM and the number observed by the LAT. Thus, high energy photon emission is associated with some but not all of the brighter GBM events. Deeper studies of the properties of the small population of high energy emitting bursts may eventually yield a better understanding of these entire phenomena.
\end{abstract}


\section{Introduction}
Most gamma-ray burst (GRB) detections have been limited to photons below a few MeV. High energy emission above 100 MeV was only detected a few times by the EGRET instrument (Dingus et al. 1995) and more recently by AGILE (Giuliani et al. 2008). The $Fermi$ Gamma-ray Space Telescope, launched in 2008, has enlarged the opportunity to study high energy emission from GRBs by providing a large aperture and wide field of view. There are two instruments onboard the $Fermi$ satellite: the Gamma-ray Burst Monitor (GBM; Meegan et al. 2009) and the Large Area Telescope (LAT; Atwood et al. 2009). These detectors overlap energy bands to span from 8 KeV to above 100 GeV. At the current moment, there are several GRB instruments in orbit that can provide significantly more precise localizations (e.g. Swift; Gehrels et al. 2004, INTEGRAL; Winkler et al. 2003, AGILE; Giuliani et al. 2008). This combination provides a unique opportunity to study the physical mechanisms of GRBs across the entire electromagnetic spectrum.

The $Fermi$/LAT covers a photon energy range from below 20 MeV to more than 300 GeV with an effective field-of-view (FOV) $\sim$ 2.4 sr (Atwood et al. 2009) while the second $Fermi$ instrument, the GBM, is sensitive to the range from 8 keV to 40 MeV. The GBM covers the entire sky except that occulted by the Earth. Realtime GRB triggers are generated by the GBM (Meegan et al. 2009) at a rate of $\sim$ 250 events per year (Paciesas et al. 2010). However, of those events simultaneously observed by the LAT, only 27 have been detected at a threshold of more than $\sim$ 10 high energy photons\footnote{http://fermi.gsfc.nasa.gov/ssc/observations/types/grbs/grb$\_$table/}, corresponding to a rate of $\sim$ 9 GRBs per year (Granot 2010).

Using the matched filter technique, the $Fermi$ detection threshold was reduced to a level of $\sim$ 6 high energy photons
with the concomitant identification of 4 additional detections as reported by Akerlof et al. (2010, 2011, hereafter A10, A11) and Zheng et al. (2012b, hereafter Z12), namely GRB 080905A, 091208B 090228A, and 081006A. In Section 2, we present the high energy photon detection of three new GRBs using this matched filter technique. We discuss the photon count distribution of these events and their correlations at lower energies in Section 4 and summarize our conclusions in Section 5.

\section{Detection of GRB 110709A, 111117A and 120107A using the matched filter technique}
Following the method developed by Akerlof et al. in A10 and A11, we have searched the GRB database generated by the GBM and $Swift$ from the end of the period covered by our earlier papers, A10, A11 and Z12, up to January 15, 2012.
%The end date is set to be Aug. 6 2011 because after that, the LAT data has been updated from Pass 6 to Pass 7 and Pass 6 data is no longer available. Since the change between Pass 6 and Pass 7 data is still not fully understand, we remain using the LAT Pass 6 data as before to perform the searching.
For the GBM, the catalog\footnote{http://heasarc.gsfc.nasa.gov/W3Browse/fermi/fermigtrig.html} extraction start date is July 10, 2010 following the period studied by A10 and Z12 while for $Swift$, the comparable catalog\footnote{http://swift.gsfc.nasa.gov/docs/swift/archive/grb$\_$table.html/} extraction started at April 1, 2010, following the period reported in A11. These two periods correspond to 554 and 654 days during which the respective instruments recorded 348 and 149 GRB events. For the following analysis, the data can be divided into two sets: events triggered by the GBM alone and events triggered by $Swift$ with or without the GBM. Since the angular resolution of $Swift$ is considerably better, any additional information from the GBM provides no further constraints. Events of interest must be visible to the LAT which imposes a boresight angle restriction of less than 74$^\circ$. 191 GBM and 43 $Swift$ triggers pass this LAT acceptance criterion including 15 GBM/$Swift$ simultaneous detections. For the GBM-tagged events which impose looser restrictions on the burst location, we attempted to set a fluence lower limit of 5 $\mu$ergs/cm$^2$. Consistent with recommendations from the Fermi Science Support Center, the LAT photon directions with respect to the zenith were required to be less than 105$^\circ$. The selection of high energy GRB candidates presented in this paper is based on the criterion that the number of random background events with greater or equal matched filter weight within the appropriate trigger classification be less than 0.05 events (ie. 5\% false positive probability).

The matched filter weight prescriptions for the GBM and $Swift$-triggered data are given in papers A10 and A11 respectively. The principle difference is that the poorer angular resolution of the GBM demands an additional clustering function. The cumulative distributions for random events drawn in green in Figures 2 and 3 of the two respective papers are shown in Figure 1 along with arrows that indicate the values of the matched filter weights for the three GRB events described below. The actual time duration of an event does not directly modify the matched filter weight algorithm - the analysis uses a standard time window of 47.5 s in all cases.

Note that there are two changes to the LAT data format during the search period covered by this paper. First of all, class 4 (ultraclean photon) has been officially assigned to photons since April 2011. Since this is a proper subset of class 3 (clean), we treat class 4 photons the same as class 3 in this analysis. Secondly, the LAT data stream processing was switched from Pass 6 to Pass 7 after August 6, 2011. Since Pass 6 data is no longer available, we are restricted to Pass 7 data following this date.

\begin{figure}[!]
%\begin{figure}[!hbp]
\centering
   \includegraphics[width=.48\textwidth]{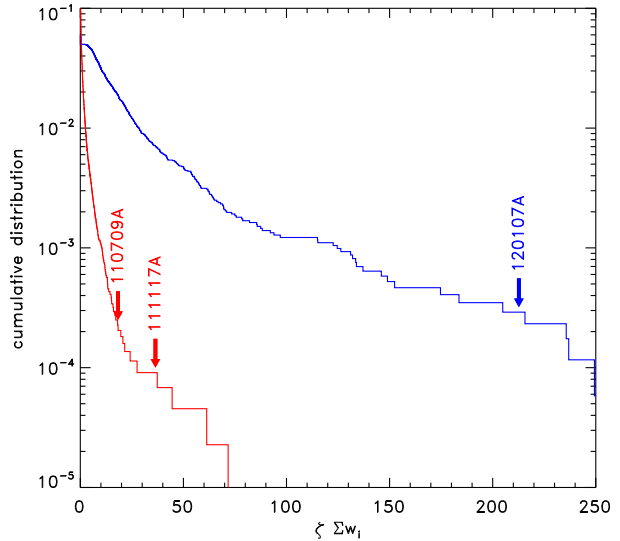}
   \caption{Cumulative distributions for matched filter weights for GBM and $Swift$ triggers for random background fields. The GBM triggers are shown in blue and the $Swift$ triggers are shown in red. This plot depicts data taken from Figures 2 and 3 of papers A10 and A11. \label{ran_wt}}
\end{figure}

\subsection{GRB 120107A detected from the GBM catalog}
GRB 120107A is a long burst trigger detected by the GBM (trigger 347620337; McBreen 2012) at 09:12:15 UT with T$_{90}$ = 23 s. The initial GBM trigger\footnote{This information was obtained electronically via Internet Socket protocol from GCN.} gives RA = 223.10$^\circ$, Dec = -77.96$^\circ$ with an uncertainty of $\sim$ 7$^\circ$. Our pipeline localized the GRB from the LAT data (Zheng \& Akerlof 2012a) to RA = 246.40$^\circ$, Dec = -69.93$^\circ$ with an uncertainty of $\sim$ 0.5$^\circ$, $\sim$ 10$^\circ$ from the GBM estimate. 8 high-energy photons in LAT data above 100 MeV were found to be associated with the GRB with the highest energy of 1.78 GeV at 8.1 s after the trigger and a total of 2 photons with energy greater than 1 GeV. The matched filter weight value is ${\zeta}{\sum}w_i$ = 212.704. The random background probability for such a value is  $2.9 \times 10^{-4}$ corrsponding to an estimated background of 0.019 events over the set of GBM triggers.

One concern for this detection is an observing direction from the LAT that was close to the Earth's limb with zenith angle $\sim 100^\circ$. At such locations, there is possible photon contamination arising from charged particle interactions with the Earth's atmosphere. As recommended by the Fermi Science Support Center, a photon zenith angle cut of less than $105^\circ$ was maintained. To address any remaining doubts, an additional check was performed to test if atmospheric interactions could mimic a GRB signal. 957 random fields were obtained at similar boresight and zenith angles. None of these 957 random fields have a matched weight value exceeding GRB 120107A (${\zeta}{\sum}w_i$ = 212.704) and the maximum matched weight for the ensemble was ${\zeta}{\sum}w_i$ = 20.3. Thus the false positive probability for detection of GRB 120107A is less than 0.1\%.

\begin{deluxetable}{crrrcr}
 \tabcolsep 0.4mm
 \tablewidth{0pt}
 \tablecaption{~ List of high energy photons in LAT data}\label{tab:Tab_Highweight}
  \tablehead{\colhead{$i^a$} & \colhead{$t^b$} & \colhead{$\theta$ ($^\circ$)$^c$} & \colhead{$E$ (MeV) $^d$ } & \colhead{$c^e$} & \colhead{$w_i ^f$}}
\startdata
\\
\multicolumn{6}{c}{GRB 110709A~~~~~${\zeta}{\sum}w_i$ = 18.3}\\
\\
 1 &  13.407 &  1.842 &  111.886 & 2  &     8.1016  \\
 2 &  29.033 &  0.225 &  166.984 & 2  &     4.3833  \\
 3 &  40.147 &  1.422 &  422.933 & 2  &     3.5022  \\
 4 &  40.156 &  3.453 &  155.548 & 3  &     1.8198  \\
 5 &  37.525 &  5.910 &  164.091 & 4  &     0.6086  \\
 6 &  20.680 &  8.278 &  105.362 & 2  &     0.5176  \\
 7 &  13.113 &  4.277 &  114.032 & 1  &     0.3313  \\
 8 &   4.555 &  6.615 &  146.916 & 1  &     0.2737  \\
 9 &  32.889 &  8.436 &  141.596 & 1  &  $<$0.0001  \\
10 &  42.104 &  2.165 & 3119.496 & 1  &  $<$0.0001  \\
\\
\multicolumn{6}{c}{After~~ 47.5 s}\\
\\
11 & 234.227 &  0.168 & 2393.739 & 3  &     9.0863  \\
12 & 121.076 &  1.172 &  112.404 & 4  &     2.1045  \\
13 & 235.321 &  0.784 &  196.466 & 4  &     1.3384  \\
14 & 119.413 &  3.096 &  222.938 & 4  &     0.5538  \\
15 &  66.932 &  5.604 &  104.551 & 2  &     0.3331  \\
16 &  96.246 &  2.881 &  149.746 & 1  &     0.0374  \\
17 &  71.234 &  3.179 &  293.516 & 1  &     0.0270  \\
18 & 245.604 &  2.288 &  131.289 & 1  &     0.0177  \\
19 & 196.575 &  5.795 &  115.638 & 2  &     0.0155  \\
20 & 106.043 &  1.806 &  856.576 & 1  &     0.0124  \\
21 & 229.291 &  3.100 &  240.405 & 1  &     0.0073  \\
22 &  95.347 &  6.989 &  123.333 & 2  &     0.0046  \\
23 & 211.107 &  4.805 &  172.602 & 1  &     0.0038  \\
24 & 180.161 &  8.913 &  102.384 & 1  &     0.0013  \\
25 & 222.976 &  5.027 &  429.738 & 2  &     0.0004  \\
26 &  84.755 &  9.907 &  192.132 & 2  &     0.0003  \\
27 & 179.455 &  3.155 &  435.976 & 2  &     0.0003  \\
28 & 218.203 &  8.347 &  139.996 & 1  &  $<$0.0001  \\
29 & 103.439 &  5.476 &  267.993 & 1  &  $<$0.0001  \\
30 &  71.445 &  7.359 &  366.642 & 1  &  $<$0.0001  \\
31 & 106.735 &  6.721 &  207.040 & 1  &  $<$0.0001  \\
32 &  80.399 &  7.336 &  484.452 & 1  &  $<$0.0001  \\
33 & 221.441 &  6.477 &  365.971 & 1  &  $<$0.0001  \\
34 &  98.638 &  8.916 &  550.749 & 1  &  $<$0.0001  \\
35 & 158.756 &  6.381 &  432.003 & 1  &  $<$0.0001  \\
\\
\hline
\\
\multicolumn{6}{c}{GRB 111117A~~~~~${\zeta}{\sum}w_i$ = 36.43}\\
\\
 1 &  1.686  & 0.865  &  364.183 & 4  & 205.9576    \\
 2 &  1.847  & 3.731  &  129.791 & 4  &  22.6106    \\
 3 &  7.722  & 5.504  &  131.676 & 1  &   0.3835    \\
 4 &  4.784  & 7.574  &  143.790 & 1  &   0.1454    \\
 5 &  5.994  & 4.317  &  203.799 & 1  &   0.0600    \\
\\
\hline
\\
\multicolumn{6}{c}{GRB 120107A~~~~~${\zeta}{\sum}w_i$ = 212.7}\\
\\
 1 &  7.262  & 0.312  &  500.525 & 4  & 411.3013    \\
 2 & 16.184  & 0.779  &  168.999 & 4  &  51.5555    \\
 3 & 24.399  & 1.221  &  371.791 & 4  &  15.3563    \\
 4 & 25.300  & 0.210  &  146.027 & 2  & *14.0574    \\
 5 &  8.111  & 0.031  & 1783.549 & 4  &   0.2759    \\
 6 & 49.199  & 0.912  & 1606.824 & 2  &   0.1747    \\
 7 &  1.181  & 5.983  &  339.577 & 4  &   0.0635    \\
 8 & 10.289  & 8.322  &  127.321 & 1  &   0.0492    \\
\enddata
\tablenotetext{a}{photon ID number}
\tablenotetext{b}{time after the trigger}
\tablenotetext{c}{distance to the new estimated GRB location}
\tablenotetext{d}{energy of each photon}
\tablenotetext{e}{photon class}
\tablenotetext{f}{weight of each photon}
\tablenotetext{*}{indicates diminished $w_E$ for highest energy triplet cluster photon}
\end{deluxetable}

\subsection{GRB 110709A detected from the $Swift$ catalog}
GRB 110709A is a long burst detected by both $Swift$ (Holland et al. 2011) and GBM (Connaughton 2011) with T$_{90}$ = 50 s. No optical afterglow was found (e.g. Xin et al. 2011). Ten high-energy LAT photons above 100 MeV are found to be associated with the GRB within our time window. The highest energy photon for GRB 110709A is 3.1 GeV at 42.1 s after the trigger. Our analysis estimates a matched weight value ${\zeta}{\sum}w_i$ = 18.3. Using the prescription described earlier, the estimated random event background is 0.0098 over the entire 43 $Swift$ trigger set.

Like many other LAT GRBs, considerable high-energy emission extends to 250 s (e.g. GRB 080825C, Abdo et al. 2009d; GRB 080916C, Abdo et al. 2009b; GRB 090510, Abdo et al. 2009c; GRB 090902B Abdo et al. 2009a). A single photon greater than 1 GeV  was also detected at 234.2 s post-trigger. However due to spacecraft slewing, the LAT boresight angle for the GRB location moved quickly from 53$^\circ$ when the burst occurred to greater than 80$^\circ$ within 100 s. At such large boresight angles, the LAT instrument sensitivity is greatly reduced and inferences about extended emission after 100 s needs to be taken with caution. We have listed the extended emission photons after 47.5 s separately from the photon within 47.5 s in Table 1.

\subsection{GRB 111117A detected from the $Swift$ catalog}
Interestingly, GRB 111117A is a short GRB detected by both $Swift$ (Mangano et al. 2011; Sakamoto et al. 2012) and GBM (Foley \& Jenke 2011) with T$_{90}$ = 0.6 s  but no reported optical afterglow. It is the only short GRB presented in this paper but similar events have previously been detected with high energy emission (e.g. GRB 090510, Abdo et al. 2009c; GRB 081024B, Abdo et al. 2010). A total of 5 high-energy photons in LAT data above 100 MeV are found to be associated with the GRB with the highest energy of 364 MeV at 1.686 s after the trigger. All 5 high energy photons are within 8 s after the trigger. Our analysis of GRB 111117A gives a matched weight value, ${\zeta}{\sum}w_i$ = 36.43 with a corresponding random background of 0.0039.

Since the burst location is relative close to the Galactic plane (gl = 122.7317$^\circ$, gb = -39.8611$^\circ$), there is a possible contamination from photons of Galactic origin. To address this issue, a similar study was performed as described for GRB 120107A. Since GRB 111117A is well localized by $Swift$, the same weight calculations were performed on 1188 LAT data sets with 47.5 s durations obtained from exactly the same celestial location during the four weeks before the event occurred. None of the 1188 trials had a matched weight value exceeding GRB 111117A (${\zeta}{\sum}w_i$ = 36.43) and the trial ensemble maximum was ${\zeta}{\sum}w_i$ = 12.73. Thus the false positive probability for detection of GRB 111117A is less than 0.084\%.

Table 1 lists all high-energy photons in the LAT data related with these three GRBs. Figure 2 depicts the LAT photon sky map and Figure 3 shows the lightcurve of the 3 GRBs.

%\clearpage
%\begin{figure}[!]
\begin{figure}[!hbp]
\centering
   \includegraphics[width=.23\textwidth]{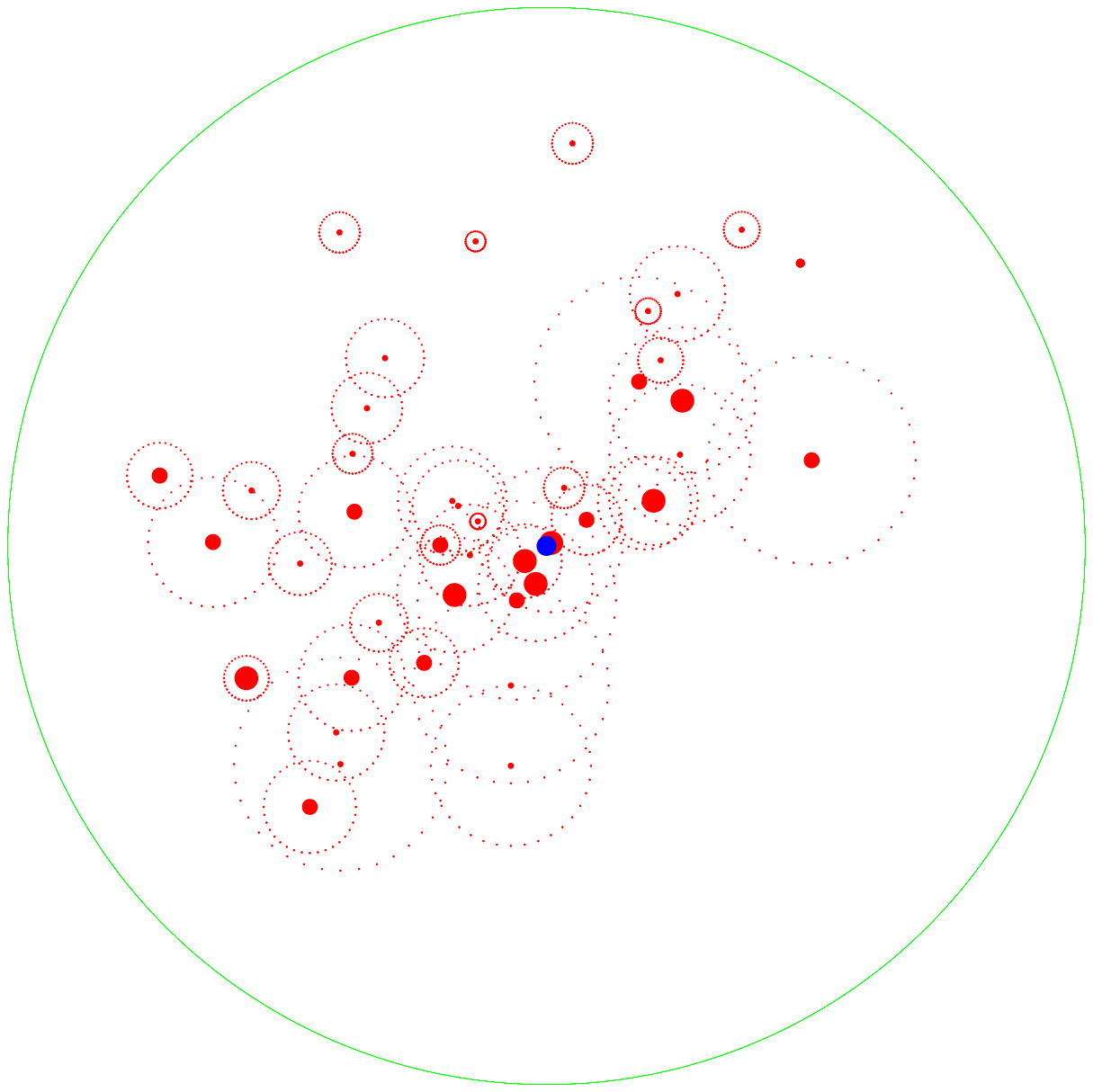}
   \includegraphics[width=.23\textwidth]{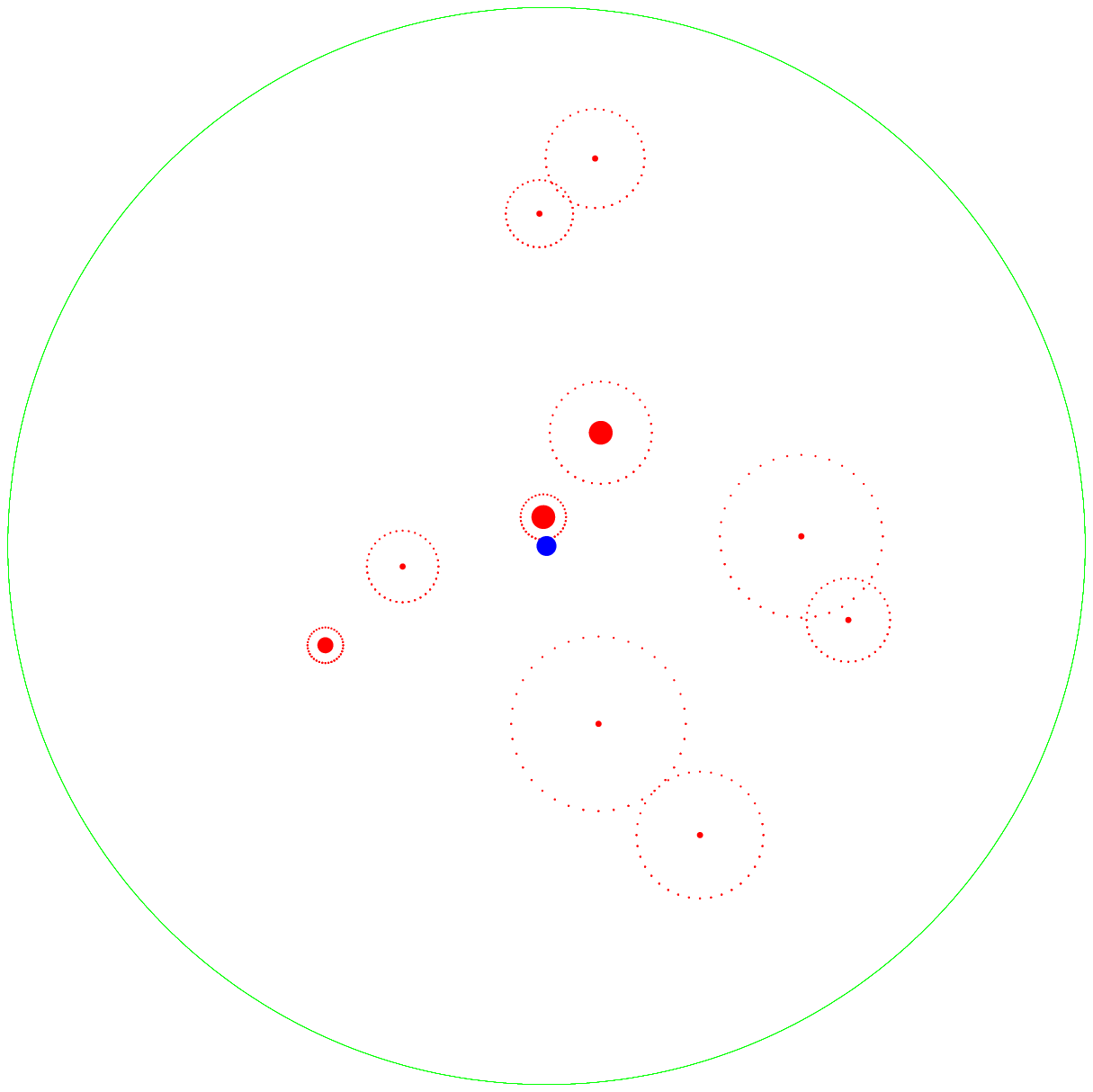}
   \includegraphics[width=.23\textwidth]{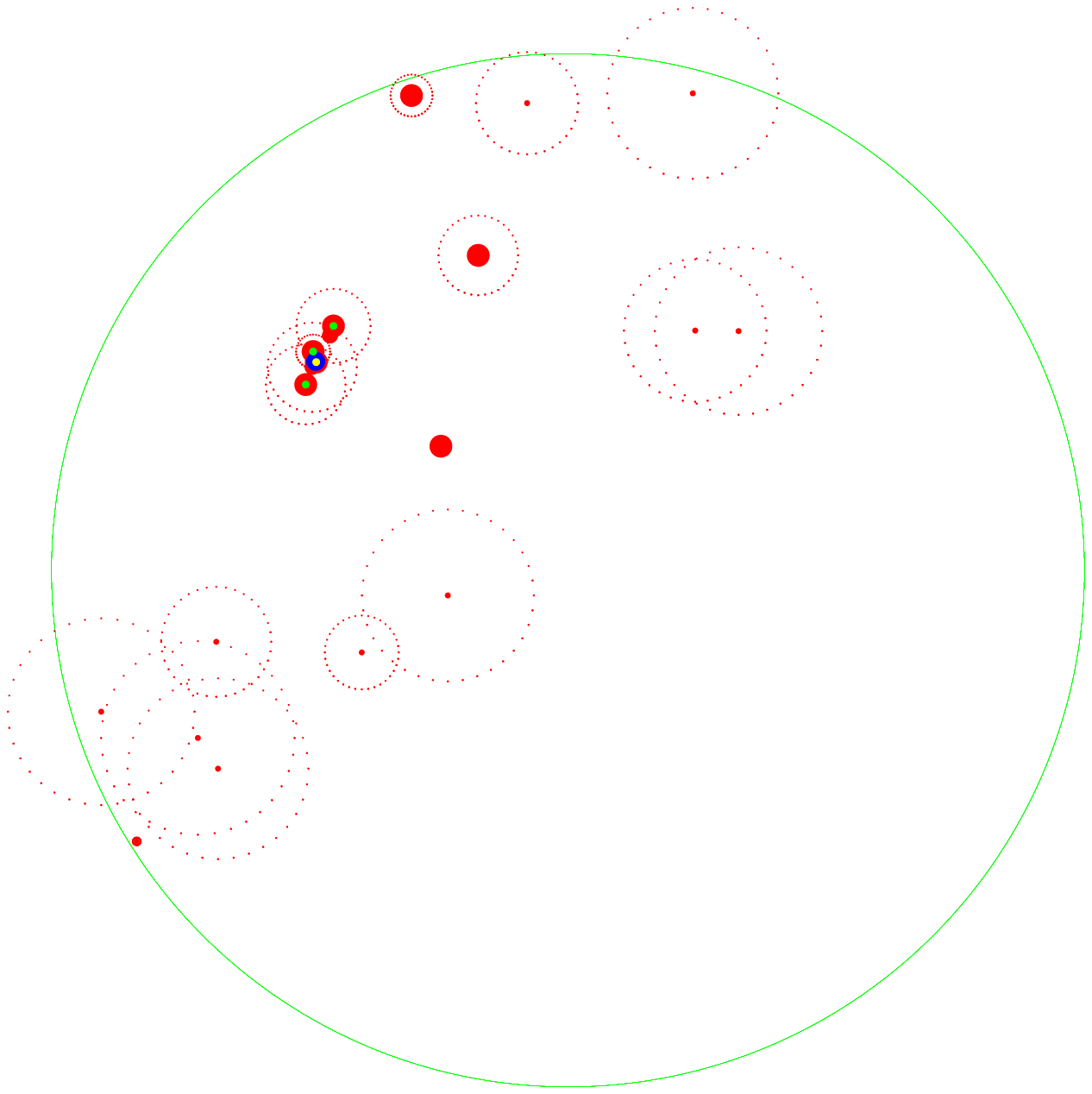}
   \caption{LAT high energy photons sky map for GRB 110709A (top left, duration is 250 s), 111117A (top right) and 120107A (bottom left). The diameter of each dot is proportional to its statistical weight. Thus, the largest diameters represent Event Class 3, etc. The dotted circles around each point indicate the $1-\sigma$ errors. The figure is centered on the nominal coordinates furnished by the GBM; the blue dot shows the GRB coordinates. The large green circle depicts the boundaries of the 16.0$^\circ$ radius cone that defines the fiducial boundaries for the cluster search. The plot axes are aligned so that North is up and East is to the right. \label{skymap}}
\end{figure}

\begin{figure}[!]
%\begin{figure}[!hbp]
\centering
   \includegraphics[width=.48\textwidth]{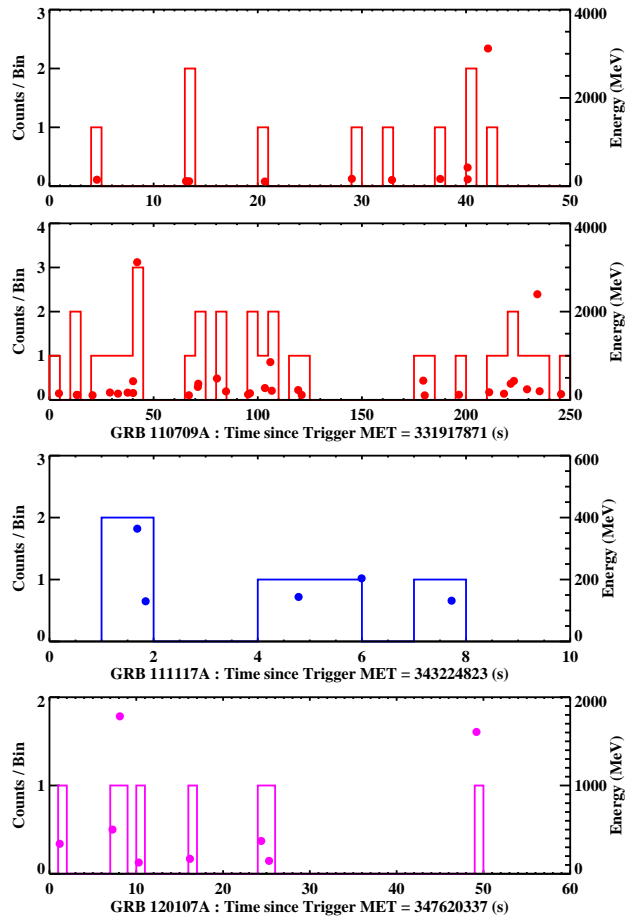}
   \caption{LAT photon light curves for GRB 110709A (1st panel for 0 - 50 s and 2rd panel for 0 - 250 s), 111117A (3th panel) and 120107A (4th panel). The filled circles represent the energy for each photon. \label{lightcurve}}
\end{figure}

\subsection{Pipeline for near realtime detection}
In the process of performing this work, a pipeline for automatically downloading and analyzing LAT data has been developed based on IDL software using the matched filter technique. The results are distributed online\footnote{http://www.rotse.net/LAT/}. This opens the possibility to detect LAT GRBs in near real time once the LAT data is available. In fact, GRB 120107A was the first LAT GRB identified by the pipeline and was announced immediately (Zheng \& Akerlof 2012a). A $Swift$/XRT ToO observation was requested and approved but unfortunately could not be performed since the GRB location was close to the Sun. As this manuscript was being completed, another GRB 120316A (Zheng \& Akerlof 2012c) was detected by our LAT data pipeline and confirmed in location by the IPN Collaboration (Hurley et al. 2012). The successful detection of GRB 120107A and 120316A by our technique provides an opportunity for guiding optical follow-up observations using the higher accuracy of the matched filter LAT localizations. Prior to the $Fermi$ launch, we expected that the LAT would provide prompt estimates of GRB locations with error circles of the order of a few degrees in radius that would enable ground-based facilities with large fields of view such as our ROTSE project to locate the optical counterparts quickly enough to enable a wide variety of observations at all wavelengths. We soon learned that the LAT data stream was constrained to infrequent downlinks so such possibilities were eliminated by the mission design. Given the low rate of GRB detections, we came to the conclusion that any increase in the number of identified events would be useful. By applying the matched filter techniques to the LAT data, we have identified 7 GRBs so far in LAT data, clearly demonstrating the advantage for detecting the fainter LAT GRBs. Although the LAT data downlink has a median latency time of $\sim$ 4-5 hours, there is still some chance for guiding optical follow-up observation. Although the probability of finding an optical counterpart is low, for brighter events (e.g. GRB 990123, Akerlof et al. 1999; GRB 080319B, Racusin et al. 2008; GRB 110205A, Zheng et al. 2011) such detections are not impossible.

\section{Discussion}
Using the matched filter technique, we have identified 7 previously unreported LAT GRBs (3 in this paper and 4 in previous publications). This adds nearly 30\% to the LAT GRB sample of 27 events found by the Fermi collaboration. Despite a significant decrease in effective threshold intensity, the number of new GRB identifications is surprisingly low. As a reference point, the brightness of local objects taken from a spatially homogeneous population should follow a power-law with index = -3/2 for the cumulative distribution. For the brightest and presumably closest GBM-detected GRBs ($\sim$15\%), the photon count distribution follows this quite closely  (see Figure 4-7 in Paciesas et al. 2012, also Figure 1 and 2 in Nava et al. 2011). Thus, a doubling of the number of LAT events might have been expected. To explore this more carefully, the LAT GRB sample has been modeled with a similar underlying power-law behavior.

%\begin{figure}[!]
\begin{figure}[!hbp]
\centering
   \includegraphics[width=.48\textwidth]{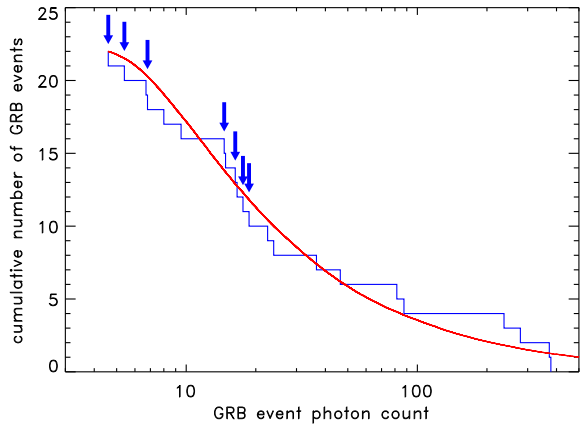}
   \caption{Cumulative distribution of the LAT GRB sample of 22 events as a function of photon number, corrected for detector effective area. The units for the horizontal axis are photons/m$^2$ for the 47.5 s interval following the GBM trigger. The actual event distribution is shown by the blue line while the Monte Carlo simulation is in red. The seven events identified by the matched filter technique are marked by arrows. \label{dist}}
\end{figure}

In all of the following, we will focus on the total number of photons detected for each GRB event within the energy band defined by the specific detector, LAT or GBM. This strategy partially reduces the redshift effects that lead to additional variability for fluence. For the LAT GRB subsample based on the matched filter technique, the time window is restricted to the 47.5 s interval following the associated GBM trigger. Twenty-two GRBs match these criteria including the 7 new identified events described here and in our earlier papers. The LAT photon counts used in the following analysis are determined by the number of quanta with weights greater than 0.1, corrected for the instrument effective area, a function of the LAT boresight angle obtained from the Fermi Science Support Center\footnote{http://www.slac.stanford.edu/exp/glast/groups/canda/archive/pass6v3/lat$\_$Performance.htm}. Table 2 lists the GRBs included in this analysis.
\begin{deluxetable}{lrrr}
 \tabcolsep 0.4mm
 \tablewidth{0pt}
 \tablecaption{~ List of 22 LAT GRB samples}\label{tab:Tab_LAT_GRBs}
  \tablehead{\colhead{Name} & \colhead{Boresight} & \colhead{Photons$^a$} & \colhead{Photons$^b$} \\
  \colhead{} & \colhead{Angle($^{\circ}$)} & \colhead{Observed} & \colhead{Corrected}}
\startdata
080825C & 60  & 10  & 36.6  \\
080916C & 49  & 125 & 279.0 \\
081024B & 19  & 12  & 16.6  \\
090217  & 34  & 14  & 22.5  \\
090510  & 13  & 176 & 237.1 \\
090902B & 50  & 164 & 378.1 \\
090926A & 47  & 177 & 372.2 \\
091003  & 12  & 11  & 14.8  \\
110120A & 15  & 7   & 9.5   \\
100325A & 8   & 5   & 6.7   \\
100414A & 70  & 7   & 87.5  \\
110428A & 34  & 5   & 8.0   \\
100724B & 51  & 10  & 23.9  \\
110721A & 42  & 25  & 46.4  \\
110731A & 3   & 61  & 81.5  \\
080905A & 30  & 3   & 4.6   \\
091208B & 56  & 5   & 14.6  \\
090228A & 16  & 5   & 6.8   \\
081006A & 16  & 12  & 16.3  \\
110709A & 54  & 7   & 18.7  \\
111117A & 14  & 4   & 5.4   \\
120107A & 56  & 6   & 17.6 
\enddata
\tablenotetext{a}{observed photons with matched weight great than 0.1,  with units of : ph (47.5s)$^{-1}$}
\tablenotetext{b}{boresight angle corrected photons, with units of : ph m$^{-2}$ (47.5s)$^{-1}$}
\end{deluxetable}

%(From Carl) With this instrumental issue resolved, our major focus is the distribution of photon number at the high energy end of the spectrum above 100 MeV as detected by the Fermi/LAT. We expect that this is a convolution of a number of features of bursts including their distribution in space, the angular distribution of photons around the central jet axis and the intrinsic intensity distribution for those events that satisfy the LAT detection energy window. To keep the discussion as simple as possible, the distributions of interest are the number of photons detected by the Fermi/LAT at energies greater than 100 MeV and the number of photons detected by the Fermi Gamma‐Burst detector (GBM) in its energy range from XXX to YYY keV. In the former case, the time window is 50 seconds following a GBM trigger.  Based on earlier investigations (Akerlof 2010), this interval captures xx\% or more of the high energy flux for most events. The equivalent total photon number estimate for the GBM is obtained by multiplying the peak photon flux by T50 as obtained from the ???. This estimate is much more accessible than going back to the original GBM data archives but by comparison of 10 events, we have established the correspondence is valid with an error of about 20\%.

First of all, the total number of events in this sample is quite small and covers a large dynamic range (roughly a factor of 100). Thus, any satisfactory fitting technique must avoid sensitivity to arbitrary data binning. We have turned to the Smirnov-Cram\'er-Von Mises (S-C-VM) statistic (Eadie et al. 1971) to achieve such independence. The input to the fitting process started with the random selection of an event direction with respect to the LAT z-axis restricted by the maximum boresight acceptance angle of 74$^\circ$. The photon number was generated by the inversion of the formula for the cumulative distribution of a power law and multiplied by the LAT effective area.
%computed by the IDL function, FEFFAREAANGLE, obtained from the Fermi Science Support Center.
For each event, the number of photons was tested against estimates of the matched filter detection efficiency computed earlier in the course of this investigation. The results are shown in Figure 4. The best fit was obtained with an exponent of -1.8 $\pm$ 0.3 for the differential distribution. For this value, the S-C-VM statistic estimated that 84\% of all samples selected from the same parent distribution would be more poorly matched. This result with a cumulative distribution exponent of -0.8 is significantly different than the -3/2 value for bright GBM events for which the S-C-VM test would predict about 4\% likelihood. The high probability that the actual data is fit by a simple power law also suggests that there is no compelling argument for dividing the sample into separate dim and bright components.

The natural next question is how many GRBs might possibly be detected with energetic photons if they originate from a homogeneous distribution. During the 1230 days that span the interval between the first and last event in Table 2, 838 GBM GRB triggers were recorded. Using the spacecraft flight information and the power law distribution of LAT photon count determined above, we find that 32.8\% or 275 events would have been detected at high energies, a factor of 12 times greater than has been obtained. Why some GRBs emit high energy photons but most do not is one of the most critical questions facing this subfield of astrophysics.

One obvious possibility is that LAT GRBs are specifically associated with only the events that are brightest at lower energies. Thus, we need to explore possible correlations with low energy data obtained simultaneously from the GBM. For reasons mentioned above, we have chosen to use an estimator of the total photon number in the GBM energy range determined by the product of T$_{50}$ and the peak  flux over a 1024 ms interval. This permits the comparison of the number of LAT photons with the number of GBM photons. For the 22 events listed in Table 2, five were omitted because their T$_{50}$ duration was less than 1 s. The remaining 17 are plotted in Figure 5. Although there is wide dispersion, a Pearson correlation coefficient of 0.537 indicates that the fluxes are correlated with a probability of 99.4\%. We have adopted the event median photon count as a robust estimator of GBM intensity. For the 18 events in Figure 5, this corresponds to a count of 205 photons in the GBM energy band. Relative to the entire GBM catalog, this is only the brightest 12.3\%. Doubling this number, we estimate that out of the 275 GBM events that were observable within the LAT field of view, only 68 would have been associated with a high enough GBM flux. This is still a factor of 3 greater than observed. Two conclusions emerge: (1), LAT GRB events are strongly associated with luminous GBM events and (2), even after accounting for the luminosity bias, a substantial fraction of GRBs still do not emit detectable high energy photons. The fact that, occasionally, short duration GRBs are also found to be high energy photon emitters suggests that this phenomenon is a generic property of energetic jets uncorrelated with the details of the progenitor. We still cannot offer any idea why the apparent photon count distribution is described by such a slowly decreasing power law.

\begin{figure}[]
%\begin{figure}[!hbp]
\centering
   \includegraphics[width=.48\textwidth]{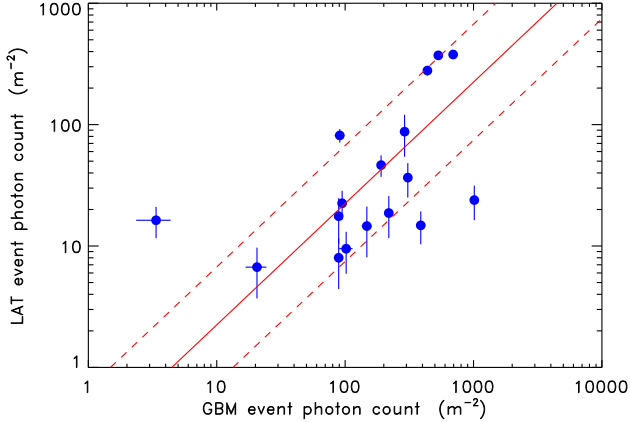}
   \caption{Scatter plot of the GBM photon count and LAT photon count for the 17 events with T$_{50}$ durations greater than 1 s taken from Table 2. The diagonal solid red line indicates the median ratio of LAT to GBM photon counts. The dashed red lines correspond to ratios 3 times larger or smaller. \label{det_corr}}
\end{figure}

\section{Summary}
We have described three new LAT GRBs identified by the matched filter technique. GRB 120107A was selected with the aid of the GBM catalog while GRB 110709A and 111117A were tagged by $Swift$. We have explored the distributions of photon counts from both GBM and LAT energy ranges. The shallow power law distribution of the numbers of photons in each event and the apparent correlation with lower energies is an intriguing development which should be tested with greater statistics. We hope that this has begun to address the mysterious nature of the relation between the very large number of GBM events and the much smaller subset at higher energies. Given the low statistics of the latter, we hope every effort will be made to encourage the most efficient search techniques of the available data stream (see also Rubtsov et al. 2012).

\acknowledgments
We thank the anonymous referee for helpful comments and suggestions for improving this manuscript.
We gratefully acknowledge the assistance of Valerie Connaughton at the Marshall Space Flight Center and Chris Shrader at the Fermi Science Support Center for their help in obtaining specific GBM data and providing information about some detailed aspects of the Fermi mission. This research is supported by the NASA grant NNX08AV63G and the NSF grant PHY-0801007.


\begin{thebibliography}{}

\bibitem[{{Abdo} {et~al.}(2009a)}]{abdo09a} {Abdo}, A.~A., et al., 2009a, ApJ, 706, L138 %090902B

\bibitem[{{Abdo} {et~al.}(2009b)}]{abdo09b} {Abdo}, A.~A., et al., 2009b, Science, 323, 1688 %080916C

\bibitem[{{Abdo} {et~al.}(2009c)}]{abdo09c} {Abdo}, A.~A., et al., 2009c, Nature, 462, 331 %090510 

\bibitem[{{Abdo} {et~al.}(2009d)}]{abdo09d} {Abdo}, A.~A., et al., 2009d, ApJ, 707, 580 %080825C

%\bibitem[{{Abdo} {et~al.}(2010)}]{abdo10} {Abdo}, A.~A., et al., 2010a, ApJS, 183, 46 %TS

\bibitem[{{Abdo} {et~al.}(2010)}]{abdo10} {Abdo}, A.~A., et al., 2010, ApJ, 712, 558 %081024B, fluence-fluence

%\bibitem[{{Ackermann} {et~al.}(2010)}]{ackermann10} {Ackermann}, M., et al., 2010, ApJ, 717, L127 %090217A

\bibitem[{{Akerlof} {et~al.}(1999)}]{akerlof99} {Akerlof}, C.~W., et al., 1999, Nature, 398, 400

\bibitem[{{Akerlof} {et~al.}(2010)}]{akerlof10} {Akerlof}, C., Zheng, W., Pandey, S.~B., McKay, T.~A., 2010, ApJ, 725, L15 (A10)

\bibitem[{{Akerlof} {et~al.}(2011)}]{akerlof11} {Akerlof}, C., Zheng, W., Pandey, S.~B., McKay, T.~A., 2011, ApJ, 726, 22 (A11)

\bibitem[{{Atwood} {et~al.}(2009)}]{atwood09} {Atwood}, W.~B., et~al., 2009, ApJ, 697, 1071

%\bibitem[{{Band} {et~al.}(1993)}] {band93} {Band}, D., et~al. 1993, ApJ, 413, 281

\bibitem{} Connaughton, V., 2011, GCN Circ., 12133

\bibitem{} Connaughton, V., 2012, private communication

\bibitem[{{Dingus } {}(1995)}] {dingus95} {Dingus}, B. L., 1995, Astrophys. Space Sci., 231, 187

\bibitem{} Eadie, W. T., Drijard, D., James, F. E., Roos, M. and Sadoulet, B., "Statistical Methods in Experimental Physics", North-Holland, 1971.

\bibitem{} Foley, S. \& Jenke, P., 2011, GCN Circ., 12573

\bibitem[{{Gehrels} {et~al.}(2004)}] {gehrels04} {Gehrels}, N., et al., 2004, ApJ, 611, 1005

\bibitem[{{Giuliani} {et~al.}(2008)}] {giuliani08} {Giuliani}, A., et al., 2008, A\&A, 491, 25

\bibitem[{{Granot} (2010)}] {granot07} Granot, J. 2010, in The Shocking Universe: Gamma-ray Bursts and High Energy Shock Phenomena,Venice, September 14-18, 2009 (Bologna: Italian Physical Society) (arXiv:1003.2452)

\bibitem{} Holland, S. et al., 2011, GCN Circ., 12118

\bibitem{} Hurley, K, et al. 2012, GCN Circ., 13073

\bibitem{} Mangano, V. et al., 2011, GCN Circ., 12559

\bibitem{} McBreen, S., 2012, GCN Circ., 12826

%\bibitem[{{Medvedev} (2000)}] {medvedev09} {Medvedev}, M.~V., 2000, ApJ, 540, 704 

\bibitem[{{Meegan} {et~al.}(2009)}] {meegan09} {Meegan}, C., et al., 2009, ApJ, 702, 791 

\bibitem{} Nava, L., et al., 2011, A\&A, 530, A21

\bibitem[{{Paciesas} {et~al.}(2010)}] {paciesas10} {Paciesas}, W., et al., 2010, BAAS, 41, 669

\bibitem[{{Paciesas} {et~al.}(2012)}] {paciesas12} {Paciesas}, W., et al., 2012, ApJS, 199,18

%\bibitem[{{van der Horst } (2008)}] {vander08} {van der Horst}, A.~J., 2008, GCN Circ., 8341 %081006A

\bibitem[]{Racusin} Racusin, J. et al., Nature, 455, 183

\bibitem{} Rubtsov, G. I., Pshirkov, M. S. \& Tinyakov, P. G., 2012, MNRAS, 421, L14

\bibitem[{{Winkler} {et~al.}(2003)}] {winkler03} {Winkler}, C., et al., 2003, A\&A, 411, L1

\bibitem{} Xin, L., et al., 2011, GCN Circ., 12125

%\bibitem[{{Zhang}(2011)}] {zhang11a} {Zhang}, B. B., et al., 2011, ApJ, 730, 141 

\bibitem{} Zheng, W., et al., 2011, arXiv:1111.0283, accepted for publication in ApJ

\bibitem{} Zheng, W. \& Akerlof, C., 2012a, GCN Circ., 12822

\bibitem[{{Zheng} {et~al.}(2012)}]{zheng12} {Zheng}, W., Akerlof, C., Pandey, S.~B., McKay, T.~A., Zhang, B.-B., Zhang, B., 2012b, ApJ, 745, 72 (Z12)

\bibitem{} Zheng, W. \& Akerlof, C., 2012c, GCN Circ., 13070

\end{thebibliography}
\end{document}